\documentclass[preprint,pra,twocolumn,10pt]{revtex4}%
\usepackage{amsfonts}
\usepackage{amsmath}
\usepackage{amssymb}
\usepackage{graphicx}%
\begin{document}
\title{Finite temperature coherence of the ideal Bose gas in an optical lattice}
\author{Gevorg Muradyan}
\author{James R. Anglin}
\affiliation{\textit{Fachbereich Physik, TechnischeUniversit\"{a}t Kaiserslautern, D-67663
Kaiserslautern, Germany}}

\begin{abstract}
In current experiments with cold quantum gases in periodic potentials,
interference fringe contrast is typically the easiest signal in which to look
for effects of non-trivial many-body dynamics. In order better to calibrate
such measurements, we analyse the background effect of thermal decoherence as
it occurs in the absence of dynamical interparticle interactions. We study the
effect of optical lattice potentials, as experimentally applied, on the
condensed fraction of a non-interacting Bose gas in local thermal equilibrium
at finite temperatures. We show that the experimentally observed decrease of
the condensate fraction in the presence of the lattice can be attributed, up
to a threshold lattice height, purely to ideal gas thermodynamics; conversely
we confirm that sharper decreases in first-order coherence observed in
stronger lattices are indeed attributable to many-body physics. Our results
also suggest that the fringe visibility 'kinks' observed in F.Gerbier et al.,
\textit{Phys. Rev. Lett.} \textbf{95}, 050404 (2005) may be explained in terms
of the competition between increasing lattice strength and increasing mean gas
density, as the gaussian profile of the red-detuned lattice lasers also
increases the effective strength of the harmonic trap.

\end{abstract}
\startpage{1}
\endpage{ }
\maketitle

\section{Introduction}

The first investigations of dilute Bose-Einstein condensates (BECs) in
periodic potentials \cite{1,2,3} launched a research field which continues to
expand steadily \cite{4,5}. The theoretical interest of the subject lies in
the wide range of analogous phenomena for which it can provide an idealized
model: spin-spin interactions in crystals \cite{6,7} and general Bose-Hubbard
Hamiltonian dynamics\cite{8}, effective magnetic fields with non-abelian gauge
potentials\cite{9,10}, and even gravitational phenomena \cite{11}. Experiments
exploiting the high tunability of optical lattices have observed the Mott
insulator phase transition \cite{12,13}, the Berezinskii-Kosterlitz-Thouless
transition\cite{14}, strong quantum depletion\cite{15}, extended coherency
with number-squeezed states\cite{16}, second-order atom tunneling\cite{17},
resonantly enhanced tunneling\cite{18}, and repulsively bound atom
pairs\cite{19}. The combination of tunable optical lattices and ultracold
temperatures provides enough control over atomic motion to investigate
numerous questions of fundamental interest and importance.

A workhorse experimental technique for such investigations is the measurement
of matter wave decoherence. Measurements of collective excitation spectra can
provide independent information about many-body dynamics \cite{12}; but such
measurements are in general somewhat more involved. The coherent population
fraction can be inferred fairly straightforwardly from measurements of fringe
contrast in matter wave interference. A drop in coherent population fraction
can signal the formation of interparticle correlations due to many-body
interactions \cite{20}. It can also be caused, however, by simple heating of
the sample. Moreover, ramping up an optical lattice alters the density of
states in a gas, and so can change its equilibrium properties even if heating
as such can be suppressed. To use coherent population fraction as a signal of
many-body physics, therefore, one must understand these background effects
well enough to subtract them accurately.

The question of temperature and critical temperature change during the process
of loading an optical lattice with atoms has already received substantial
study \cite{22,23,24,25,26,27,28,29}. Unfortunately the problem turns out to
be quite complicated in practice. While it is straightforward to show that
ramping up a uniform lattice extremely slowly will adiabatically cool a
non-interacting ultracold gas, interactions and a time-dependent trapping
potential are unavoidable experimental complications. And although experiments
are able to allow quite long times for the gas to equilibrate after the
lattice has been fully turned on, the available experimental lifetimes set
limits on how slowly the lattice potential can be ramped up, in order to reach
full lattice strength before the gas escapes the trap. Under present
conditions it therefore seems difficult to ensure that the gas really remains
in equilibrium during the entire experiment. Some loss of condensate fraction
does occur when the lattice strength is raised and lowered in a cycle, and
although it is small enough to allow ample observations of Bose-Einstein
condensates in optical lattices, it does indicate non-equilibrium entropy
generation during ramping.

The theory of heating during loading of a condensate into an optical lattice
is thus a subject of active research. It is an important task for cold atom
theory, because direct experimental measurement of very low temperatures in
highly condensed Bose gases remains technically difficult. (The standard
fitting of the thermal fraction's momentum distribution to the
Maxwell-Boltzmann formula becomes problematic when the thermal fraction is
small.) The purpose of this paper is to study the effects of lattice
potentials, as experimentally applied with focused laser beams and with
harmonic trapping potentials superimposed, on the density of states of the
ideal Bose gas. Our main contribution is to identify a basic competition
between density of states and spatial density, which occurs because of the
particular way optical lattices are implemented in current experiments, and
which considerably moderates the effect of weaker lattices on gas coherence.

As a first exercise in identifying the quantitative consequences of these
effects in experiments, we model several recent experiments under the
simplifying assumptions that the gas does remain in global thermal equilibrium
at all times, and moreover that it maintains a constant temperature. We
determine this temperature by fitting to experimental data from before the
lattice potentials are ramped up. This scenario would be expected in the case
where the trapped gas was maintained in equilibrium through scattering with a
reservoir of thermal gas, which was itself too slightly perturbed by the
time-dependent lattice to undergo any temperature change. For shallow lattices
our isothermal model turns out to fit experimental data somewhat better than
the alternative scenario of isentropic evolution optical lattices, which would
be expected if the lattice-trapped gas could be regarded as isolated, and if
the lattice were ramped up slowly enough for equilibrium still to be
maintained. Our quantitative results thus provide some somewhat surprising
input for the further non-equilibrium studies which will ultimately be
required in this problem. Moreover, they serve to confirm the dynamical
competition effect we identify, which must play an important role in
determining how the gas coherence depends on the lattice potential, regardless
of the precise statistical ensemble which should describe the gas.

The paper is organized as follows. We first indicate the parameter regime in
which an ideal Bose gas in thermal equilibrium is expected to be a good model
for experimental systems. We then describe the theory we will use for a
Bose-condensed ideal gas in an optical lattice at finite temperature. Along
with this full version of our ideal gas theory, whose predictions we will
compute numerically, we introduce some further approximations that will allow
analytical results, and help interpret our numerical curves.

We then present our numerical calculations and compare them with several sets
of experimental data, including experimental configurations in which the gas
dynamics is effectively three-, two-, or one-dimensional. In particular we
consider the fringe visibility "kinks" seen in recent experiments, where the
downward sloping curve of first order matter coherence versus increasing
lattice strength has a shoulder. We show and explain a very similar effect in
the ideal gas coherent fraction, which occurs at closely comparable lattice strengths.

In our final section we summarize our results and provide an outlook sketch.

\section{Thermal equilibrium}

We assume interaction between particles sufficient to maintain equilibrium by
Boltzmann scattering. In a truly collisionless regime, an adiabatic change of
potential would simply preserve the population in each energy level, and so
would not change the number of condensed atoms. In considering an ideal gas in
equilibrium, however, we naturally mean that we consider time scales long
enough for interactions in the form of collisions to generate the
time-averaged behavior that is represented by a grand canonical ensemble; but
we do not admit interactions strong enough to alter the instantaneous
probability distributions in that ensemble, which therefore depends only on
the non-interacting Bose gas Hamiltonian.

Cold dilute gases in experiments can certainly be made very weakly
interacting, so the experimental relevance of our model depends on the
relation between the experimental time scale $\tau_{X}$ and the timescale
$\tau_{C}$ for collisional equilibration. Quantum kinetic theories for
ultracold atoms support the identification of the latter timescale with
Boltzmann's scattering time:
\begin{equation}
\tau_{C}=\frac{1}{8\pi a^{2}\rho}\sqrt{\frac{m}{2k_{B}T}}, \label{1}%
\end{equation}
where $\rho$ is the gas density, $T$ is the gas temperature, $a$ is the s-wave
scattering length for low momentum transfer collisions, and $m$ is the
particle mass. This gives an equilibration timescale on the order of $\tau
_{C}\approx40ms$ for $^{87}Rb$ with densities of $\rho\approx10^{13}cm^{-3}$.
Since lattice ramp time scales in experiments are generally much longer than
this, our assumption of equilibrium because $\tau_{C}\ll\tau_{X}$ should be
adequate for experimental comparison.

The validity of our ideal gas approximation then depends on the ratio between
$\tau_{C}$ and the relevant time scale $\tau_{D}$ for single-particle
dynamics. What then is this $\tau_{D}$? A global definition in terms of the
rate of change of single particle energy eigenstates would be appropriate for
experiments in the collisionless kinetic regime, studying the first onset of
interaction effects. For current experiments approaching phenomena such as
quantum phase transitions due to onsite interactions, a local criterion is
more relevant. We therefore consider $\tau_{D}$ to be the time scale for
quantum tunneling between two adjacent optical lattice sites. We can invoke
the energy-time uncertainty relation to estimate this tunneling time as
$\tau_{D}=\hbar/\delta E$ for $\delta E$ the width of the lowest energy band.
At optical lattice depths of order $\ 40E_{r}$ ($E_{r}=\hbar^{2}k^{2}/2m$
being the 'recoil energy') this implies $\tau_{D}\approx100ms$. For the
strongest experimental lattices we thus have $\tau_{D}\gtrsim\tau_{C}$, so
that particles typically collide before leaving one lattice site. The ideal
gas approximation therefore breaks down, and interparticle interactions begin
to have non-trivial effects beyond merely enforcing equilibrium over long time
scales. Accurately assessing such effects is one of the major challenges in
current many body theory; our goal in this paper is simply to identify the
onset of this regime in current experiments.

For shallower optical lattices, however, the tunneling time rapidly decreases:
$\tau_{D} \approx1ms$ for a lattice depth of $12E_{r}$. With $\tau_{D}\ll
\tau_{C}\ll\tau_{X}$, particles will typically tunnel through many lattice
sites before scattering, so that the resulting equilibrium distribution of
single-particle energies will include the band structure of the lattice, but
the effects of harmonic trapping over length scales much longer than the
lattice spacing can be computed in local density approximation. This theory
should apply up to the onset of dynamically (as opposed to kinetically)
significant interactions at high lattice potentials. As we will see, it does
indeed fit experiments well up to threshold lattice strengths. This agreement
is not trivial, however, but depends upon the cancellation of two basic
features of the single-particle dynamics in typical experimental potentials.
The interaction onsets identified in this paper therefore provide useful
information for interpreting experiments on ultracold atoms in optical lattices.

\section{Model potential for optical lattice plus harmonic trap}

We consider an ideal Bose gas in a harmonic trap to which has been added a
periodic potential created by counterpropagating laser beam pairs in $x$, $y$
and $z$ directions. It turns out to be very important to represent the
experimental fact that the focused red-detuned laser beams which generate the
lattice potential have gaussian radial envelope profiles\cite{30}:
\begin{equation}
V=\sum_{i=1}^{3}\left[  \frac{m\omega_{0,i}^{2}r_{i}^{2}}{2}+V_{i}[\sin
^{2}(kr_{i})-1]e^{\frac{-2(r^{2}-r_{i}^{2})}{W_{i}^{2}}}\right]  \label{2}%
\end{equation}
Here $\omega_{0,i}$ is the angular frequency of the harmonic trapping
potential in the $i$th direction, $k$ is the wavenumber of the 1D lattice
potential formed by the AC Stark effect in the standing waves of the $i$th
direction's laser pair, and $W_{i}$ is these beams' `waist' (radius at which
laser intensity has fallen by a factor $1/e^{2}$). We write $\sin^{2}%
(kr_{i})-1$ instead of $-\cos^{2}(kr_{i})$ because we will treat differently
the smooth gaussian envelope of the lattice minima and the gaussian modulation
of the lattice barrier height.

The importance of the finite waist in the gaussian beam profile is as follows.
All of these experiments use red-detuned light, so that the laser intensity
maxima provide the lattice potential minima. This means that the AC Stark
potential felt by the atoms is equivalent to a broad gaussian well, plus a
repulsive lattice whose height diminishes with radius (according to the
gaussian envelope). The spatial extent of the gas cloud is limited by the
combined slowly-varying potential which is the superposition of the confining
harmonic trap, plus the broad gaussian envelope of the attractive laser
intensity maxima. Since the beam waist is in our cases much wider than the
extent of the gas cloud in the trap, we can Taylor-expand the Gaussian profile
of the beam, so that the envelope is approximately quadratic over the extent
of the cloud. We must therefore consider that strengthening the lattice
potential simultaneously strengthens the harmonic trapping confinement of the gas.

As well as the lattice potential minima becoming less deep towards the edges
of the cloud, the height of the lattice barriers also decreases with radius;
and this changes the local density of states for the atoms. For the
experimental cases considered in this paper, however, the variation in the
local width of the first energy band over the extent of the gas cloud is never
greater than about $5\%$. This difference becomes even smaller for the
stronger lattices that induce significant decoherence. We will therefore
simplify our calculations considerably, at small cost in accuracy, by
neglecting the effectively small local variation in lattice barrier height.
The reason we can neglect the beam profile for the lattice barriers, but must
include it in the attractive potential envelope, is that the relevant energy
scale for the confining effect of the lattice minimum envelope is the gas
chemical potential $\mu$, while the energy scale relevant to the effect of
lattice height variation on bandwidth is at least $E_{r}$, which is much
higher than $\mu$ in the cases we consider. We will therefore represent the
potential induced by the red-detuned laser light as a superposition of a
smooth harmonic well and a uniform repulsive lattice:
\begin{equation}
V\rightarrow\sum_{i}\left[  V_{i}\sin^{2}(kr_{i})+\frac{m\omega_{i}^{2}%
r_{i}^{2}}{2}\right]  \label{3}%
\end{equation}
where%
\begin{equation}
\omega_{i}^{2}=\omega_{0,i}^{2}+\frac{4}{mW_{j}^{2}}V_{j}+\frac{4}{mW_{l}^{2}%
}V_{l} \label{4}%
\end{equation}
for $i,j,l$ in cyclic order.

As a further technical simplification with negligible impact on the accuracy
of our results, we will then replace the actual sinusoidal lattice with a
parabolic spline potential, which is piecewise quadratic, and in particular is
a periodic succession of upward and downward parabolas (`biparabolic'). As
Fig.1 shows, this approximation is very accurate, but it will allow us to
compute band structures more easily by taking advantage of analytical
properties of confluent hypergeometric functions, instead of dealing with
Mathieu functions. We will therefore finally write
\begin{equation}
V\rightarrow\sum_{i}\left[  V_{i}U(kr_{i})+\frac{m\omega_{i}^{2}r_{i}^{2}}%
{2}\right]  \;,\label{5}%
\end{equation}
having defined the periodic function
\begin{equation}
U(\xi)\equiv\frac{1-(-1)^{n}}{2}+(-1)^{n}\frac{2}{\pi^{2}}(\xi-n\pi
)^{2},\label{6}%
\end{equation}
where the integer $n$ is chosen for given $\xi$ such that $(n-1/2)\pi
\leqslant\xi\leqslant(n+1/2)\pi$.%
\begin{figure}
[ptb]
\begin{center}
\includegraphics[
height=1.8273in,
width=2.943in
]%
{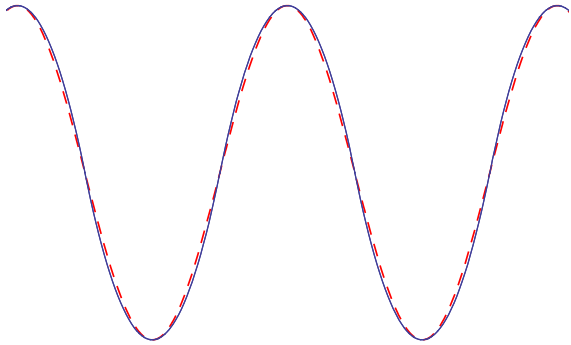}%
\caption{Biparabolic periodic potential (solid line) used in this paper, in
place of the actual sinusoidal lattice (dotted line).}%
\end{center}
\end{figure}

For the uniform biparabolic lattice, (\textit{i.e.} with $\omega_{i}$ set to
zero in $V$ above), the single particle Schr\"{o}dinger equation factorizes
and the dispersion relation between energy and quasimomentum $\vec{p}$ is
given by a some of one-dimensional energies,
\[
E(\vec{p})=\sum_{i=1}^{3}E_{i}(p_{i})\;.
\]
We show in the Appendix that the $E_{i}(p_{i})$ can be found implicitly from
dimensionless equations of the form
\begin{equation}
\cos\left(  \frac{\pi p_{i}}{\hbar k}\right)  =1-G\left(  \frac{E_{i}}{E_{r}%
},\frac{V_{i}}{E_{r}}\right)  \;,\label{7}%
\end{equation}
where $G$ is a particular combination of confluent hypergeometric functions
\cite{31} (see the Appendix). These equations can be solved numerically to
produce an explicit numerical interpolating function for $E(\vec{p})$ in the
uniform lattice, for any given lattice strengths $V_{i}$.

\section{Thermal equilibrium in local density approximation}

We assume that the state of the (dynamically) non-interacting gas may be
represented with a grand canonical ensemble, and this determines the number of
non-condensed atoms $N_{NC}$ as a function of chemical potential $\mu$ and
temperature $T$. We also take note of the fact that the gas density, locally
averaged over a range of lattice cells, varies very slowly on the lattice
scale. This allows us to use a local density approximation, as follows. We
will consider
\begin{equation}
\label{mu-r}\tilde{\mu}(\vec{r})\equiv\mu-\sum_{i=1}^{3}\frac{m\omega_{i}%
^{2}r_{i}^{2}}{2}%
\end{equation}
for fixed $\vec{r}$ to be the `local chemical potential', and compute the
grand canonical density of non-condensed atoms $\rho(\vec{r})$ in the uniform
lattice at temperature $T$ and chemical potential $\tilde{\mu}(\vec{r})$. We
will then integrate over $\vec{r}$ to obtain the total $N_{NC}$. In the next
Section we will then compare the theoretical $N_{NC}$ to the experimentally
measured total particle number $N$, for various recent experiments. The
difference $N_{C}=N-N_{NC}$ is inferred as the number of ideally
Bose-condensed atoms. The coherency fraction $N_{C}/N$ will then be
interpreted as the coherency fraction that would be expected in those
experiments if only ideal gas physics were involved.

We therefore write
\begin{equation}
\rho(\vec{r})=\frac{1}{(2\pi\hbar)^{3}}\int\!\frac{d^{3}p}{\exp\left[
\beta(E(\vec{p})-\tilde{\mu}(\vec{r}))\right]  -1},
\end{equation}
where $1/\beta\equiv k_{B}T$. We perform the $\vec{p}$ integrals using the
so-called extended-zone scheme, integrating over quasimomenta within the
successive energy bands. In most cases we will be at low enough temperatures
to restrict our integration to the first energy band only.

To now integrate over $\vec{r}$, we expand the Bose-Einstein denominator in a
rapidly converging Taylor series to obtain
\begin{align}
N_{NC}  &  =\int\!d^{3}r\,\rho\nonumber\\
&  =\sum_{n=1}^{\infty}\int\frac{d^{3}p}{(2\pi\hbar)^{3}}e^{-n\beta E(\vec
{p})}\int\!d^{3}r\,e^{n\beta\widetilde{\mu}(\vec{r})}\nonumber\\
&  =C\sum_{n=1}^{\infty}\frac{e^{n\beta\mu}}{n^{3/2}}\int\,d^{3}p\,e^{-n\beta
E(\vec{p})} \label{8}%
\end{align}
for
\[
C\equiv\left(  \frac{k_{B}T}{2\pi m\hbar^{2}}\right)  ^{\frac{3}{2}}\frac
{1}{\omega_{x}\omega_{y}\omega_{z}}%
\]
and, since we are always considering samples in which some condensate is
present,
\[
\mu\rightarrow\sum_{i=1}^{3}E_{i}(0)\;.
\]
We can perform the $\vec{p}$ integrals in the last line of Eqn.~(\ref{8})
numerically, for each $n$, using the $E(\vec{p})$ determined by numerically
solving (\ref{7}). The sum converges rapidly, and so we obtain a numerical
figure for $N_{NC}$ for any given temperature $T<T_{c}$ and set of lattice
strengths $V_{i}$. Eqn.~(\ref{8}), evaluated in this way, is the main result
of this paper. In combination with experimental measurements of total particle
number, and temperature estimates, it will provide the 'theoretical' curves
plotted against experimental data in the next Section.

\subsection{Further approximations}

Since our main result (\ref{8}) is evaluated numerically, it is useful to
compare it with simpler approximations whose physical content is more
transparent. We will now present several of these, of which two will also be
displayed, as dashed curves, in some of the next Section's plots.

As the periodic potential becomes stronger, the right-hand-side (RHS) of
Eq.(\ref{7}) approaches a linear function of $E_{i}/E_{r}$ (within each energy
band). As an approximation that should become valid for high lattice barriers,
therefore, we take the RHS to be an exactly linear function and obtain the
following dispersion relation for the first energy band:
\begin{align}
E_{i}(p_{i})  &  =\frac{E_{i}^{\max}+E_{i}^{\min}}{2}-\frac{E_{i}^{\max}%
-E_{i}^{\min}}{2}\cos\left[  \frac{\pi p_{i}}{\hbar k}\right] \nonumber\\
&  \equiv\bar{E}_{i}-\Delta_{i}\cos\left[  \frac{\pi p_{i}}{\hbar k}\right]
\;. \label{9}%
\end{align}
Here $E^{\min}$ and $E^{\max}$ are the first energy band's boundary values.
Although we determine these band edges from numerical calculations for our
model lattice potential, this approximation for $E_{i}$ is of the same form
obtained in the well known tight binding approximation. We will therefore
refer to this as the tight-binding approximation for $E(\vec{p})$. Using it,
the $\vec{p}$ integral in (\ref{8}) can be performed analytically over the
first band, yielding
\begin{equation}
N_{NC}^{TB}\rightarrow(2\hbar k)^{3}C\sum_{n=1}^{\infty}n^{-3/2}\prod
_{i=1}^{3}e^{-{n\beta\Delta_{i}}}I_{0}\left(  n\beta\Delta_{i}\right)  ,
\label{111}%
\end{equation}
where $I_{0}$ is the modified Bessel function. Eqn.~(\ref{111}), referred to
as 'the tight-binding approximation' for $N_{NC}$, will be used to generate
the dashed curves in several Figures of the next Section.

For readers not intuitively familiar with modified Bessel functions, of
course, this approximated expression is no more transparent than the more
accurate numerical results of Eqn.~(\ref{8}). To indicate the dependence on
the energy bandwidth more explicitly, however, we can simplify Eqn.~(\ref{111}%
) further in the strong lattice limit $\Delta_{i}\ll k_{B}T$, where the
band\textit{width} shrinks far below the temperature, but the temperature in
turn remains well below the interband \textit{gap}. In this limit we
approximate the modified Bessel function as $I_{0}(0)=1$ and obtain
\begin{equation}
N_{NC}^{TB}\approx Li_{3/2}\left(  e^{-{\beta}%
{\textstyle\sum\limits_{i=1}^{3}}
{\Delta_{i}}}\right)  \prod_{i=1}^{3}\left[  \frac{2}{\hbar\omega_{i}}%
\sqrt{\frac{k_{B}TE_{r}}{\pi}}\right]  , \label{TBA}%
\end{equation}
where $Li_{s}\left(  \xi\right)  =\sum_{k=1}^{\infty}\xi^{k}/k^{s}$ is the
polylogarithm function.

This simple formula shows well the qualitative impact of the lattice
parameters on the noncondensed particle number. Strengthening the periodic
potential compresses the energy bands, raising the density of states and thus
increasing the noncondensed fraction. Increasing the harmonic trapping
strength instead favors condensation, and hence tends to remove non-condensed
atoms. Since the beam focus means that strengthening the lattice also
strengthens the effective harmonic trap, these two effects compete as the
lattice strength is ramped up.

This simple competition can also be seen, still using the tight-binding
dispersion relation (\ref{9}), but in the opposite temperture regime
$\Delta_{i}>>k_{B}T$. \ Using the asymptotic form $I_{0}(\xi)\approx e^{\xi
}/\sqrt{2\pi\xi}$ we obtain
\begin{align}
N_{NC}  &  =\zeta(3)\prod_{i=1}^{3}\left[  \frac{k_{B}T}{\hbar\omega_{i}}%
\sqrt{\frac{2E_{r}}{\pi^{2}\Delta_{i}}}\right] \label{11}\\
&  \equiv N_{NC}^{HO}\prod_{i=1}^{3}\sqrt{\frac{2E_{r}}{\pi^{2}\Delta_{i}}%
}\;,\nonumber
\end{align}
$N_{NC}^{HO}$ being the number of non-condensed bosons in a harmonic trap of
frequencies $\omega_{i}$ at temperature $T$ \cite{32}. ($\zeta(z)$ is the
Riemann zeta function, and $\zeta(3)\doteq1.202$). Neither of the two
simplified formulas (\ref{TBA}) and (\ref{11}) will be compared with data in
this paper, or for that matter referred to again at all; but they serve to
illustrate qualitatively the competition between trap and band compression
that is exhibited in the tight-binding approximation (\ref{12}) and in the
full theory of (\ref{8}).

In the regime of temperatures much below the bandwidth, we expect most atoms
to explore only the parabolic region of the dispersion relation, near its
minimum. This immediately suggests, as an alternative to tight binding, the
parabolic or effective mass approximation, in which the motion of particles in
periodic structures is described as motion of free particle but with an
effective mass (which is in general anisotropic):
\[
\frac{1}{m_{i}^{\ast}}=\frac{\partial^{2}E_{i}}{\partial p_{i}^{2}}%
\rfloor_{\vec{p}=0}\;.
\]
By Taylor expanding and keeping the first two terms on both sides of our
dispersion relation Eq.(\ref{7}), we obtain
\[
m_{i}^{\ast}=\frac{2m}{\pi^{2}}\frac{\partial\ }{\partial\epsilon}%
G(\epsilon,V_{i}/E_{r})\rfloor_{\epsilon=E_{i}(0)/E_{r}}\;.
\]
Now approximating $E_{i}\rightarrow m_{i}^{\ast}p_{i}^{2}/2$ and integrating
over the infinite range of $\vec{p}$, we find
\begin{align}
N_{NC}^{EM}  &  =\zeta(3)\prod_{i=1}^{3}\left[  \frac{k_{B}T}{\hbar\omega_{i}%
}\sqrt{\frac{m_{i}^{\ast}}{m}}\right] \label{12}\\
&  \equiv N_{NC}^{HO}\frac{\sqrt{m_{x}^{\ast}m_{y}^{\ast}m_{z}^{\ast}}%
}{m^{3/2}}.\nonumber
\end{align}
Referred to as 'the effective mass approximation', this simplified formula
will provide a second dashed curve for comparison in some of the Figures of
the next Section.

Thus for low temperatures as well as for higher ones, and whether we assume
tight binding or quasi-free motion, we see that the non-condensed particle
number has on one hand a tendency to increase (at the expense of the
condensate) as the lattice barrier height rises, because this raises the
effective mass and compresses the energy bands. On the other hand the
non-condensed number also tends to shrink, so that the condensate grows, when
the harmonic confining potential is strengthened. And since experiments with
focused red-detuned laser lattices make the barrier height and trapping
potential rise and fall together, though not in direct proportion, a
non-trivial competition is to be expected.

For temperatures within the bandwidth $\Delta_{i}$, the non-condensed particle
number clearly depends on the details of the dispersion relation, and no
simple approximation to Eqn.~(\ref{8}) will be very accurate. Our full
numerical results for Eqn.~(8), however, will still represent essentially this
same competition between barrier height and containment.
\begin{figure}
[ptb]
\begin{center}
\includegraphics[
height=1.8645in,
width=2.9585in
]%
{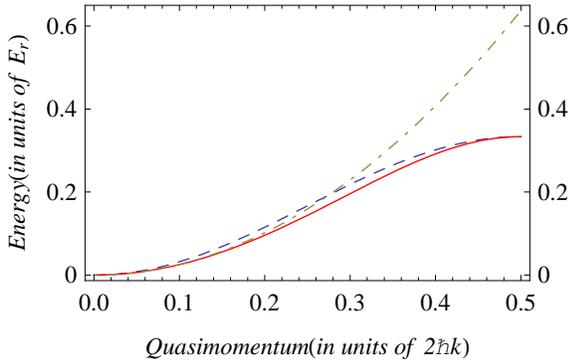}%
\caption{First band energy-quasimomentum relation for the biparabolic lattice
(solid line), in one dimension, with $V_{i}=4E_{r}$. The dashed and dotdashed
lines give the same dependance in the tight binding and effective mass
approximations, respectively. }%
\end{center}
\end{figure}

Before comparing our results with several recent experiments, we pause to show
that the competition between confinement and the barrier lattice is indeed
important. The effectively increased harmonic confinement provided by a
stronger red-detuned focused lattice can compensate to a considerable degree
for the increased barrier height, and maintain a much higher condensate
fraction than would survive if the total harmonic confinement were truly held
constant. If we have a gas of $N=1.5\ast10^{5}$ atoms in a harmonic trap with
$\omega_{x,0},\omega_{y,0}$,$\omega_{z,0}=2\pi\times\{20,20,20\}$ $Hz$, and
$T/T_{C}^{0}=0.3(15nK)$ is assumed to remain constant, then the condensate
vanishes entirely at an optical lattice height of $14E_{r}$. Including the
increased effective confinement, due to a gaussian beam profile with $1/e^{2}$
radii of 120 $\mu$m, changes the behavior completely: the condensate fraction
$f_{c}=(N-N_{NC})/N$ changes by less than ten percent over this range of
lattice heights. See Figure 3.%
\begin{figure}
[ptb]
\begin{center}
\includegraphics[
height=2.034in,
width=2.9585in
]%
{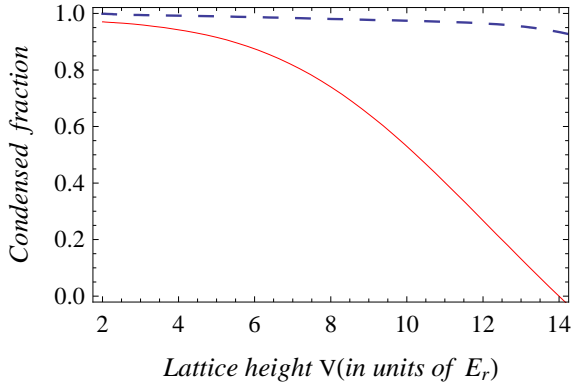}%
\caption{Condensed fraction as a function of optical lattice height with
(dashed) and without (solid) inclusion of the Gaussian profile of the beams.}%
\end{center}
\end{figure}

\section{Comparison with observations}

Figures. 4 through 7 show the results of our full numerical theory described
above(isothermal) (Eq.(\ref{8})), plotted as a solid curve (red online), as
well as of the two semi-analytic approximations (Eq.(\ref{111}) for tight
binding approximation and Eq.(\ref{12}) for effective mass approximation) and
of the isentropic evolution, when their inclusion is instructive. Superimposed
for comparison are data points directly from the experimental publications. We
take the experimentally measured total numbers $N$ of trapped particles,
together with our theoretical values for non-condensed particle numbers
$N_{NC}$, to compute the coherent fraction $f=1-N_{NC}/N$. The experimental
values for coherent fraction were measured from interference fringe constrast.

We take all parameters directly from the published values for the experiments,
with the exception of the gas temperatures. In the regime of very high
condensate fraction, quantum gas thermometry is \textit{experimentally} very
difficult, and only a rough upper bound on the experimental temperature can be
estimated ($T/T_{C}^{0}\leq0.36$, for example, in \cite{13}). We have
therefore taken temperature as a fitting parameter, tuning it so that our
theoretical curve hits the first experimental data point (\textit{i.e.} the
one for lowest lattice strength). This may be considered an application of
`coherence thermometry': we use ideal gas thermodynamics in the lattice to
infer the gas temperature from interference contrast measurements. We assume
that the gas temperature remains constant, within the same experimental data
set, for all other barrier heights. As already mentioned, this isothermal
assumption is rather crude, and could certainly be improved upon by
considering detailed quantum kinetic studies of the effects of changing
lattice strength. The fact that interpretation of the data in line with our
isothermal theory curves is nevertheless surprisingly straightforward is a
significant point, therefore, even if it should turn out to be co-incidental,
because it bears on the sensitivity of this type of experiment to quantum gas
kinetics. We will discuss this issue further in our final section.

The various experiments we consider span the range of effective
dimensionalities. In Fig.4 is shown the case where the lattice heights $V_{i}$
in all 3 directions are kept equal as they are ramped up, so that the gas is
always effectively three-dimensional. The data for this case are taken from
\cite{13}. Our fitted temperature is $T=45nK=0.29T_{C}^{0}$, where $T_{C}^{0}$
is the critical temperature for BEC in the (rather elongated) experimental
trap, which had $\{\omega_{x,0},\omega_{y,0}$,$\omega_{z,0}\}=2\pi
\times\{20,120,120\}$ $Hz$.

In Fig.~4 we see that in the 3D geometry the noninteracting, isothermal gas's
coherent fraction changes very little for lattice heights from $2E_{r}$ to
$4E_{r}$. For stronger lattices than this, it monotonically decreases, until
about $10E_{r}$; but at still higher lattice strengths, the coherence of the
non-interacting gas actually increases slightly. This is because the stronger
harmonic confinement of the gas is slightly over-compensating for the
compression of the density of states in the stronger lattice. As one would
expect, the tight-binding approximation approaches the full theory quite
closely for strong lattices. The effective mass approximation becomes quite
poor for stronger lattices, because it ignores the large band gap which is
opening up. Since it includes thermal population of within-gap modes that do
not actually exist, it greatly exaggerates the non-condensed population and
hence underestimates the coherent fraction.

The theoretical curve for the isothermal ideal gas has an excellent
qualitative and quantitative agreement with experimental data for lattice
heights up to $\sim8E_{r}$, but thereafter the observed coherence of the real,
interacting gas drops sharply. The fair agreement between the observations and
the ideal gas in effective mass approximation is clearly co-incidental, since
the drop in the effective mass approximation curve at this point is an error
in its representation of the ideal gas. As the lattice becomes stronger,
tunneling rates between lattice sites drop sharply, so that interactions
become more and more important in comparison with kinetic energies, while the
atomic interactions also become gradually greater as the gas becomes more
tightly confined within each site. The ideal gas approximation gradually
worsens, then breaks down abruptly at the Mott transition.

For this experimental case we have also computed the ideal gas coherent
fraction assuming isentropic evolution, using a straightforward but
computationally much more demanding version of our local density
approximation. (Since entropy is an extensive variable, holding it constant is
a non-local constraint.) As comparison between the solid and dotted curves of
Fig.~4 shows, the distinction between isothermal and isentropic evolution is
not dramatic for lattice strengths below those at which the ideal gas model
itself breaks down. In detail, however, the isentropic dotted curve seems to
do somewhat more poorly than the isothermal solid curve in fitting the
'shoulder' or 'kink' feature of the data, which although it is a small effect,
is common to all currently available experiments. (This 'coherency saturation
kink' will be discussed further below.) This fact would seem to be worth
noting for future non-equilibrium studies of dilute Bose gases in
time-dependent lattices.%

\begin{figure}
[ptb]
\begin{center}
\includegraphics[
height=2.2554in,
width=2.9585in
]%
{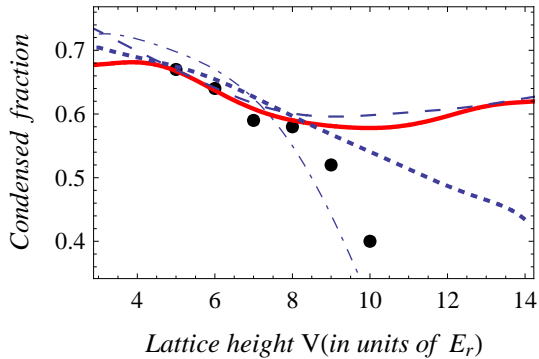}%
\caption{Coherent fraction of ideal Bose gas (solid line) as a function of
optical lattice height $V_{i}=V$, with all three lattice components equal in
strength. The dashed and dot-dashed lines show the results under the
tight-binding and parabolic approximations described in the text. The dotted
line describes isentropic evolution. Points are experimental data taken from
[13].}%
\end{center}
\end{figure}

In Fig.5 we compare our ideal gas theory with experimental data reported in
\cite{33}. In this case an effectively two dimensional situation was realized,
by ramping up one component of the lattice to the rather high value of
$8E_{r}$, and subsequently raising the lattice components in the other two
orthogonal directions together, keeping them equal. Here our fitting happens
again to find $T=45nK=0.29T_{C}^{0}$, while all the other parameters are as in
\cite{33}. Comparison shows that the ideal gas model still works very well for
the same lattice height range of 2 to 8 $E_{r}$. The effective mass
approximation is poor in this case because the lattice is always quite strong
in one direction. A mixed approximation would presumably be better, assuming
tight binding in one direction and a parabolic spectrum in the other two.%
\begin{figure}
[ptb]
\begin{center}
\includegraphics[
height=2.2978in,
width=2.9585in
]%
{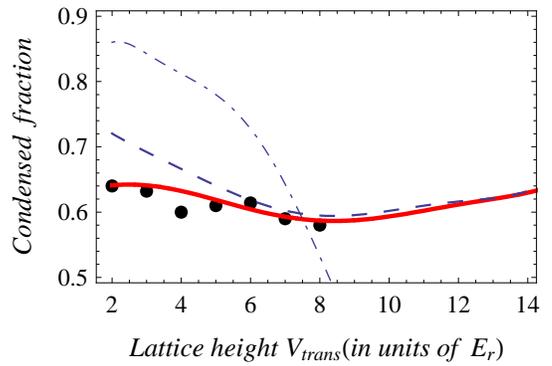}%
\caption{Coherent fraction of the ideal Bose gas (solid line) as a function of
transverse optical lattice height $V_{trans}$. The lattice strength along the
long axis of the sample is held constant at $V_{ax}=8E_{r}$. The dashed and
dot-dashed lines give the same quantity under tight-binding and parabolic
approximations, respectively. Points are experimental data taken from [32]. }%
\end{center}
\end{figure}

A quasi-one-dimensional version of this problem has also been realized
\cite{34}, and is shown in Fig.~6. The trapped gas was in this case loaded
into a strong two-dimensional lattice $40E_{r}$ high, and a weaker lattice
potential along the axial direction $V_{ax}$ was raised to various much lower
heights. In this case we see excellent agreement between the isothermal ideal
gas theory and the experimental data, albeit in a regime where the variation
is not dramatic in either one. To fit the initial experimental point, however,
we have had to take a quite high temperature, nearly twice what the initial
critical temperature would be in the same harmonic trap without the lattice:
$T=193nK=1.94T_{C}^{0}$. (This harmonic trap had $\{\omega_{x,0},\omega_{y,0}%
$,$\omega_{z,0}\}=2\pi\times\{8.7,90,90\}$ $Hz$.) With the strong transverse
lattice and initial axial lattice at $V_{ax}=2E_{r}$, however, the ideal gas
critical temperature for BEC is actually $250nK$, so our fitted temperature
was still well below the actual $T_{c}$. It is high enough, though, that the
thermal occupation of the second axial band is non-negligible, and so in this
case we have included it in our full theory curve.

The temperature of the initial condensate in this experiment, before loading
it into the lattice, was reported to be only around $50nK$. It is unclear,
therefore, whether heating during this loading process has really raised the
temperature by a factor of four, or whether many-body effects are already
suppressing coherence at weak axial lattices because of the tight transverse
confinement, or whether the high gas compression of the transverse lattice has
simply raised the rate of scattering among atoms during the time-of-flight
imaging process enough to degrade the observed interference patterns. The
simplest explanation of Fig.~6, which is that turning on the strong transverse
lattice had indeed considerably heated the samples, is by no means extremely
implausible, however. The evidence from coherence for quantum many-body
effects, as opposed to ideal gas thermodynamics, would seem to be weaker in
the one-dimensional geometry than in the higher dimensionality cases.%

\begin{figure}
[ptb]
\begin{center}
\includegraphics[
height=2.034in,
width=2.9585in
]%
{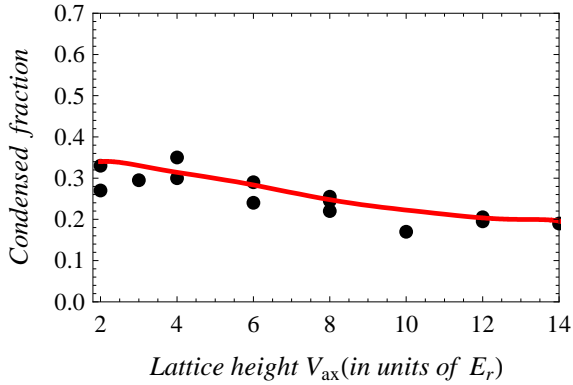}%
\caption{ Coherent fraction of the ideal Bose gas (solid line) as a function
of axial optical lattice height $V_{ax}$. The lattice height in the other two
directions is kept fixed at $V_{trans}=40E_{r}$. Points are experimental data
taken from [33].}%
\end{center}
\end{figure}

\subsection{Coherency saturation}

In all three cases discussed above, the initially falling theoretical curves
level out, and then start to increase at some value of the potential height.
This apparently counter-intuitive behavior of the non-interacting gas is in
fact easily explained, as being due to the co-incidental effective
strengthening of harmonic confinement, along with strengthening of the lattice
barriers, when the focused red-detuned laser intensities are raised. A
stronger lattice means a narrower energy band, hence a greater density of
states and larger non-condensed population, hence a lower coherent fraction.
But a tighter trap at fixed (critical) chemical potential holds fewer
non-condensed particles, and hence favors higher coherence. Which of these
competing effects dominates depends upon the lattice and trap strengths and
shapes. It also depends upon temperature. For temperatures ranging from
$T=30-60nK$ (which is $T\simeq0.1-0.2T_{C}^{0}$ for the elongated trap of
\cite{35}, with $\{\omega_{x,0},\omega_{y,0}$,$\omega_{z,0}\}=2\pi
\times\{20,200,200\}$ $Hz$), the lattice height at which the local coherency
minimum occurs ranges over $V\sim14-8E_{r}$. (Higher temperature brings the
minimum to lower lattice heights.) Even for the ideal gas the quantitative
details in this effect are involved, but the two competing factors are
qualitatively simple and clear.

It seems worth bearing them in mind when investigating observations, as in
\cite{35}, of `kinks' in the coherence fraction curve for the real gases \~{N}
lattice heights at which the steady fall of coherent fraction seems to pause,
before resuming. The fact that the coherency does resume falling, instead of
rising as in the ideal case, is evidently due to interparticle interactions.
And the true explanation of these coherency `kink' observations may indeed
have to do with the shifting of superfluid and Mott insulator shells with
lattice height in the presence of the harmonic potential. But as Fig.~7 shows,
when the number of atoms is $3.6\times10^{5}$ and the temperature is tuned to
$T=30nK$, the ideal gas coherency minimum is at $14E_{r}$, the same location
seen experimentally. The fact that only mild re-tunings of the temperature are
needed to make the ideal gas coherence curve show similar kinks at exactly the
same lattice heights indicates that coherence enhancement because of rising
effective harmonic confinement may also be playing a significant role at this
point in the experimental system. It is beyond the scope of this paper,
however, to assess the impact of this effect beyond the ideal gas regime.%
\begin{figure}
[ptb]
\begin{center}
\includegraphics[
height=6.0166cm,
width=7.831cm
]%
{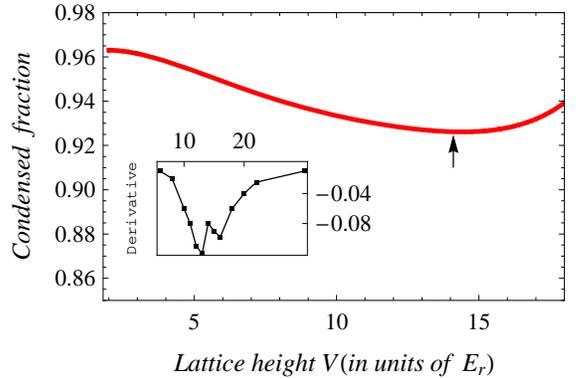}%
\caption{Coherent fraction as a function of optical lattice height under
experimental conditions of \cite{34}. The arrow shows the point where the
energy spectrum change due to harmonic trapping starts to dominate the change
due to the periodic potential. The inset is data taken from \cite{34}, and
gives a numerical derivative of the experimentally measured visibilty curve as
a function of lattice height.}%
\end{center}
\end{figure}

\section{Conclusion and outlook}

In this paper we have shown that optical lattices formed from red-detuned
gaussian profile laser beams contribute an effective harmonic confinement to a
trapped quantum gas, and that this effect needs to be taken into account in
computing the coherence fraction of an ideal Bose gas in such a lattice at a
given sub-critical temperature. It is in principle quite possible to remove
this effect by altering the magnetic trap while raising the laser intensity,
so as to hold the total harmonic confinement constant while the lattice is
raised. We have shown that doing so would significantly simplify the
interpretation of experimental results.

The good agreement between experiment and ideal gas theory that we have found
up to threshold lattice heights supports the interpretation that many-body
effects are slight up to that point, but that true many-body effects such as
number squeezing and even the Mott transition do then appear, because ideal
gas physics does not seem able to account naturally for the rather sharp drop
in coherence observed at these points. It is important to note that the
mildness of the variation with lattice height of the ideal gas's coherent
fraction is largely due to cancellation between the lattice's band compression
and the concomitant increase in harmonic trapping strength, as we have just
summarized. Without this coincidental compensation through effective harmonic
confinement, our Fig.~3 showed clearly that band compression at constant
temperature can sharply lower the coherence of an ideal Bose gas in a strong
optical lattice. It is therefore important to consider the effects discussed
in this paper when inferring many body physics from coherency data.

A weak point in our theory as it stands is certainly the assumption that the
temperature does not vary with lattice height. If the lattice is ramped on
slowly enough for the cloud to remain in global equilibrium, and if the gas
temperature is fixed by evaporation and heating processes that are insensitive
to the lattice barrier height, then the isothermal assumption may well be
valid. Determining precisely where observations depart from equilibrium ideal
gas theory may also require more careful consideration, however. It seems
quite plausible that our isothermal equilibrium curves are to some extent
mimicking the more realistic combination of adiabatic cooling with
non-equilibrium entropy generation. And perhaps, near the threshold lattice
height, adiabatic cooling of the gas is able to compensate for some
decoherence due to incipient number squeezing. If that were so, many body
physics could actually be emerging somewhat before the experimental coherence
fell below our isothermal theory curves. Or perhaps heating does increase
before quantum many body effects really predominate, so that the first fall in
coherence involves only ideal gas dynamics, and the onset of many-body
dynamics comes only later. These issues will require further study, even just
within ideal gas dynamics. Extension into weakly interacting dynamics, with
mean field theory and Bogoliubov quasiparticles, is then an obviously
desirable next step.

Finally, crude though its application in this paper has been, our temperature
fitting procedure has raised the prospect of applying coherence thermometry in
experiments on cold bosons in optical lattices, in regimes where normal
time-of-flight thermometry becomes imprecise. Better theoretical calibration,
taking into account the non-trivial kinetics issues described above, may make
this a precise technique for future experiments.

\section{Appendix}

In this Appendix we derive the transcendental equation from which the
dispersion relation in the uniform lattice may be determined numerically.

The one-dimensional time-independent Schr\"{o}dinger equation for the
biparabolic periodic potential may be written
\begin{equation}
\psi^{\prime\prime}(\xi)=(\lambda U(\xi)-\epsilon)\psi(\xi)\tag{A1}\label{A1}%
\end{equation}
where
\begin{align}
\epsilon &  =\frac{E_{i}}{4E_{r}}\nonumber\\
\lambda &  =\frac{V_{i}}{4E_{r}}\nonumber\\
\xi &  =2kx_{i}\nonumber
\end{align}
and $U(\xi)$ is given by Eq.~\ref{6}, which we re-write here:
\begin{equation}
U(\xi)\equiv\frac{1-(-1)^{n}}{2}+(-1)^{n}\frac{2}{\pi^{2}}(\xi-n\pi
)^{2}\;.\nonumber
\end{equation}
Linearly independent solutions of the equation \ref{A1} in the region
$\pi/2<\xi<3\pi/2$ are%

\begin{equation}
\varphi_{1}(\xi)=\exp\left(  -\frac{\sqrt{\frac{2\lambda}{\pi^{2}}}(\xi
-\pi)^{2}}{2}\right)  \Phi\left(  \alpha,\frac{1}{2};\sqrt{\frac{2\lambda}%
{\pi^{2}}}(\xi-\pi)^{2}\right)  ,\tag{A2}\label{A2}%
\end{equation}%
\begin{align}
\varphi_{2}(\xi) &  =(\xi-\pi)\exp\left(  -\frac{\sqrt{\frac{2\lambda}{\pi
^{2}}}(\xi-\pi)^{2}}{2}\right)  \tag{A3}\label{A3}\\
&  \times\Phi\left(  \alpha+\frac{1}{2},\frac{3}{2};\sqrt{\frac{2\lambda}%
{\pi^{2}}}(\xi-\pi)^{2}\right)  ,\nonumber
\end{align}
where $\Phi(X)$ is a confluent hypergeometric function,
\[
\alpha=\frac{1}{4}\left(  1-\frac{\epsilon}{\sqrt{\frac{2\lambda}{\pi^{2}}}%
}\right)  ,
\]
and thus the wavefunction in this region can be written as%
\begin{equation}
\Psi_{I}=c_{1}\varphi_{1}(\xi)+c_{2}\varphi_{2}(\xi),\tag{A4}\label{A4}%
\end{equation}
whith unknown constant coefficients $c_{1}$ and $c_{2}$.

Correspondingly the linearly independent solutions in the "barrier-type"
region $3\pi/2<\xi<5\pi/2$ are%
\begin{align}
\widetilde{\varphi}_{1}(\xi)  &  =\exp\left(  \frac{i\sqrt{\frac{2\lambda}%
{\pi^{2}}}(\xi-2\pi)^{2}}{2}\right) \tag{A5}\label{A5}\\
&  \times\Phi\left(  \beta,\frac{1}{2};-i\sqrt{\frac{2\lambda}{\pi^{2}}}%
(\xi-2\pi)^{2}\right)  ,\nonumber
\end{align}%
\begin{align}
\widetilde{\varphi}_{2}(\xi)  &  =(\xi-2\pi)^{2}\exp\left(  \frac{i\sqrt
{\frac{2\lambda}{\pi^{2}}}(\xi-2\pi)^{2}}{2}\right) \tag{A6}\label{A6}\\
&  \times\Phi\left(  \beta+\frac{1}{2},\frac{3}{2};-i\sqrt{\frac{2\lambda}%
{\pi^{2}}}(\xi-2\pi)^{2}\right)  ,\nonumber
\end{align}
where%
\[
\beta=\frac{1}{4}\left(  1-i\frac{\epsilon-\lambda}{\sqrt{\frac{2\lambda}%
{\pi^{2}}}}\right)  ,
\]
and the wavefunction is%
\begin{equation}
\Psi_{II}=\widetilde{c}_{1}\widetilde{\varphi}_{1}(\xi)+\widetilde{c}%
_{2}\widetilde{\varphi}_{2}(\xi), \tag{A7}\label{A7}%
\end{equation}
with unknown constant coefficients $\widetilde{c}_{1}$ and \ $\widetilde
{c}_{2}$.

Using the Bloch theorem \cite{36} \ we can write the wavefunction in the
region $5\pi/2<\xi<7\pi/2$ as\
\begin{equation}
\Psi_{III}=\exp\left[  \frac{i\pi p}{\hbar k}\right]  \Psi_{I}. \tag{A8}%
\label{A8}%
\end{equation}

Now implementing the continuity requirements of the wavefunctions and their
derivatives at the boundary points $3\pi/2$ and $5\pi/2,$ we get a homogeneous
system for the unknown coefficients, and from the requirement of nontrivial
solution we obtain the dispersion relation Eq.(7), with%
\begin{align}
G\left(  \frac{E}{E_{r}},\frac{V}{E_{r}}\right)   &  =\left(  \varphi_{1}%
(\xi)\widetilde{\varphi}_{1}^{\prime}(\xi)+\widetilde{\varphi}_{1}(\xi
)\varphi_{1}^{\prime}(\xi)\right)  \tag{A9}\label{A9}\\
&  \times\left(  \varphi_{2}(\xi)\widetilde{\varphi}_{2}^{\prime}%
(\xi)+\widetilde{\varphi}_{2}(\xi)\varphi_{2}^{\prime}(\xi)\right)
\vert_{\xi=\pi/2},\text{ \ \ \ }\nonumber
\end{align}
with prime denoting the first order derivative.

This work was supported by the Alexander von Humboldt Foundation Georg Forster
programme and The Marie Curie RTN EMALI (MRTN-CT-2006-035369).

\end{document}